\documentclass[aps,prl,showpacs,floatfix]{revtex4}

\usepackage{psfig} \usepackage{epsfig} \usepackage{amssymb}
\usepackage{amsmath}

\def\R{\mathbb{R}}

\def\<{\langle} 
\def\>{\rangle}
\begin{document}

\title{String Method for the Study of Rare Events}

\author{Weinan E$^1$, Weiqing Ren$^2$, and Eric Vanden-Eijnden$^2$}
\affiliation{$^1$Department of Mathematics and PACM,
  Princeton University, Princeton, NJ 08544\\
  $^2$Courant Institute, New York University, New York, NY 10012 }


\begin{abstract}
  We present a new and efficient method for computing the transition
  pathways, free energy barriers, and transition rates in complex
  systems with relatively smooth energy landscapes. The method
  proceeds by evolving strings, i.e.  smooth curves with intrinsic
  parametrization whose dynamics takes them to the most probable
  transition path between two metastable regions in the configuration
  space.  Free energy barriers and transition rates can then be
  determined by standard umbrella sampling technique around the
  string.  Applications to Lennard-Jones cluster rearrangement and
  thermally induced switching of a magnetic film are presented.
\end{abstract}

\maketitle

\twocolumngrid

The dynamics of complex systems are often driven by rare but important
events (for a review see e.g. \cite{hatabo90}). Well-known examples
include nucleation events during phase transition, conformational
changes in macromolecules, and chemical reactions. The long time scale
associated with these rare events is a consequence of the disparity
between the effective thermal energy and typical energy barrier of the
systems.  The dynamics proceeds by long waiting periods around
metastable states followed by sudden jumps from one state to another.

Sophisticated numerical techniques have been developed to find the
transition pathways and transition rates between metastable states in
complex systems for which the mechanism of transition is not known
beforehand \cite{mep0,TPS,NEB}. With the exception of the transition
path sampling technique \cite{TPS}, most of these methods seem to
require that the energy landscape be relatively smooth.  One typical
example of such techniques is the nudged elastic band (NEB) method
\cite{NEB}. NEB connects the initial and the final states by a chain
of states. The states move in a force field which is the combination
of the normal component of the potential force and the tangential
component of the spring force connecting the states. The spring force
helps to evenly space the states along the chain.

In this paper we propose an alternative approach for computing
transition pathways, free energy barriers, and transition rates. We
sample the configuration space with strings, i.e. smooth curves with
intrinsic parametrization such as arclength, or energy weighted
arclength which connect two metastable states (or regions), $A$ and
$B$. The string satisfies a differential equation which by
construction guarantees that it evolves to the most probable
transition pathway connecting $A$ and $B$.  One can then perform an
umbrella sampling of the equilibrium distribution of the system in the
hyperplanes normal to the string and thereby determine free energy
barriers and transition rates.

Consider the example of a system modeled by
\begin{equation}
  \label{eq:SDEV}
  \gamma \dot q = -\nabla V(q) + \xi(t).
\end{equation}
where $\gamma$ is the friction coefficient, $\xi(t)$ is a white-noise
with $\<\xi_j(t)\xi_k(0)\> = 2 \gamma k_B T \delta_{jk}\delta(t)$.
The metastable states are localized around the minima of the potential
$V(q)$. Assuming $V(q)$ has at least two minima, $A$ and $B$, we look
for the minimal energy paths (MEPs) connecting these states. By
definition, a MEP is a smooth curve $\varphi^\star$ connecting $A$ and
$B$ which satisfies
\begin{equation}
  \label{eq:MEP}
  (\nabla V)^\perp (\varphi^\star) =0,
\end{equation}
where $(\nabla V)^\perp$ is the component of $\nabla V$ normal to
$\varphi^\star$.  The MEPs are the most probable transition pathways
for (\ref{eq:SDEV}) since with exponentially high probability it is by
these paths that the system switches back and forth between the states
$A$ and $B$ under the action of a small thermal noise \cite{frwe92}.
It is interesting to note that the solutions of (\ref{eq:MEP}) also
provide relevant information about the Langevin equation
\begin{equation}
  \label{eq:SDEV2}
  \begin{cases}
    \dot q = p, \\
    \dot p = -\nabla V(q) -\gamma p + \xi(t).
  \end{cases}
\end{equation}
Indeed, the metastable regions for (\ref{eq:SDEV}) and
(\ref{eq:SDEV2}) coincide, and the transition pathways for
(\ref{eq:SDEV2}) can be easily determined from the transition pathways
for (\ref{eq:SDEV}) because they traverse the same sequence of
critical points. As a result the transition rates for (\ref{eq:SDEV2})
for an arbitrary friction coefficient $\gamma$ can be obtained by
considering the high friction evolution equation (\ref{eq:SDEV}) --
see (\ref{eq:rate}) below.


Let $\varphi$ be a string (but not necessarily a MEP) connecting $A$
and $B$.  A simple method to find the MEP is to evolve $\varphi$
according to
\begin{equation}
  \label{eq:un}
  u^\perp = - (\nabla V)^\perp (\varphi),
\end{equation}
where $u^\perp$ denotes the normal velocity of $\varphi$, since
stationary solutions of (\ref{eq:un}) satisfy (\ref{eq:MEP}). For
numerical purposes it is convenient to have a parametrized version of
(\ref{eq:un}), keeping in mind however that the parametrization can be
arbitrarily chosen since both (\ref{eq:MEP}) and (\ref{eq:un}) are
intrinsic. Denote by $\varphi(\alpha,t)$ the instantaneous position of
the string, where $\alpha$ is some suitable parametrization. Then we
can rewrite (\ref{eq:un}) as
\begin{equation}
  \label{eq:string}
  \varphi_t = -(\nabla V (\varphi))^\perp+ r \hat t,
\end{equation}
where for convenience we renormalized time $t/\gamma \to t$, $(\nabla
V)^\perp = \nabla V - (\nabla V \cdot \hat t) \hat t$, and $\hat t$ is
the unit tangent vector along $\varphi$, $\hat t =
\varphi_\alpha/|\varphi_\alpha|$. The scalar field $r\equiv
r(\alpha,t)$ is a Lagrange multiplier uniquely determined by the
choice of parametrization. The simplest example is to parametrize
$\varphi$ by arclength normalized so that $\alpha=0$ at $A$,
$\alpha=1$ at $B$.  Then (\ref{eq:string}) must be supplemented by the
constraint
\begin{equation}
  \label{eq:constraintpara}
  (|\varphi_\alpha|)_\alpha =0.
\end{equation}
which determines $r$ \cite{rem_r}. Other parametrizations can be
straightforwardly implemented by modifying the constraint
(\ref{eq:constraintpara}). For instance, a parametrization by energy
weighted arclength which increases resolution at the transition states
is achieved using the constraint
$(f(V(\varphi))|\varphi_\alpha|)_\alpha =0$, where $f(z)$ is some
suitable monitor function satisfying $f'(z)<0$. In addition, the end
points of the string need not be fixed and other boundary conditions
can be used.
%

\begin{figure}
  \center \includegraphics[width=7cm]{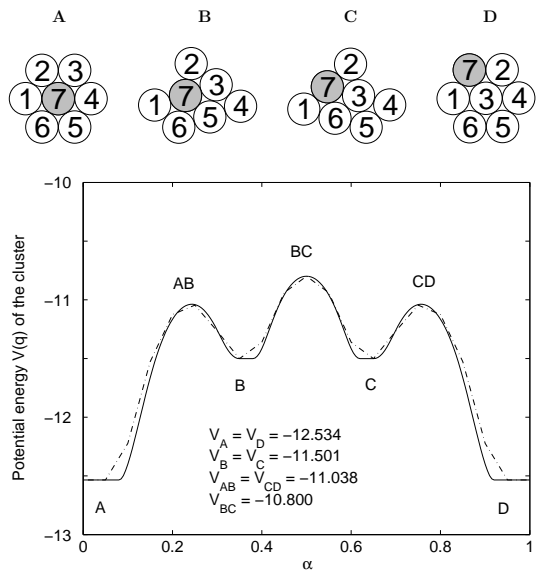}
  \caption{\label{fig:sevenatom} 
    Top figure: A transition pathway by which the central atom migrate
    to the surface in a seven-atom hexagonal Lennard Jones cluster in
    the plane.  The pictures show successive configuration
    corresponding to local minima of potential energy along the path.
    Bottom figure: the potential energy along the path in natural
    units.  The full line corresponds to a simulation with $N=200$
    discretization points along the string, and the dashed line,
    $N=20$.}
\end{figure}

Because of the intrinsic description of the string, it is very simple
to implement an efficient algorithm which solves (\ref{eq:string})
using a time-splitting type of scheme. The string is discretized into
a number of points which move according to the first term, $-(\nabla V
(\varphi))^\perp$, at the right hand-side of (\ref{eq:string}). After
a number of steps depending on the accuracy for the constraint
(\ref{eq:constraintpara}), a reparametrization step is applied to
reinforce~(\ref{eq:constraintpara}).  This costs $O(N)$ operations
where $N$ is the number of discretization points along the string. At
the reparametrization step it is also convenient to change $N$
according to the accuracy requirement for the representation of the
string.

The method certainly bears some similarities with NEB since one can
think of the introduction of the spring force in NEB as a way of
ensuring equal-distance parametrization by a penalty method -- NEB
gives an evolution equation which, in the continuum limit, is similar
to (\ref{eq:string}) but with $r$ given by $r=\kappa
\varphi_{\alpha\alpha}\cdot \hat t$ where $\kappa$ is the artificial
spring constant. As in other penalty methods, this numerical procedure
introduces stiffness into the problem if the penalization parameter,
here the elastic constant $\kappa$, is large and this limits the size
of the time step.  By using an intrinsic description, we eliminate
this problem and speed up convergence.  Furthermore we gain the
ability of using other parametrizations in a simple and flexible way.
Finally, there is no simple way to change the number of discretization
points along the chain in NEB.

It is natural to ask how the string method compares to NEB in terms of
performance. However such a comparison does not seem straightforward
since it depends on the criteria. This is discussed in detail in
\cite{ereva01}.

\begin{figure}[t]
  \center \includegraphics[width=7cm]{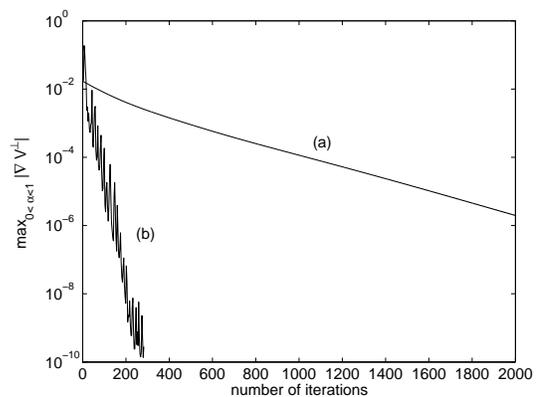}
  \caption{\label{fig:sevenatomconv} 
    The convergence history of the steepest descent method [(a)] and a
    limited memory version of Broyden's method [(b)] applied to the
    seven-atom cluster problem with $N=200$ discretization points.  We
    use $\max_{0\le\alpha\le1} |(\nabla V)^\perp|$ to measure
    accuracy.}
\end{figure}

As a first example we look at the dynamics of seven atoms interacting
via Lennard-Jones potential on the plane.  This example has been
studied in detail in \cite{deboch98}. In equilibrium the seven atoms
form an hexagon. We are interested in the process in which the atom at
the center migrates to an external position. The MEPs are not unique
for this problem. One example of MEP obtained via the string method is
shown in figure~\ref{fig:sevenatom}; the critical points along this
path coincide with the ones obtained in \cite{deboch98} by transition
path sampling.

The string equation (\ref{eq:string}) essentially amounts to finding
the MEPs by the method of steepest descent except that we are not
working with any explicit object function to minimize. We can use
advanced numerical techniques for solving nonlinear equations
\cite{kel95} to accelerate convergence to the MEP. We have developed a
limited memory version of Broyden's method where (\ref{eq:string}) is
replaced by
\begin{equation}
  \label{eq:string2}
  \varphi_t = -G^\perp (\nabla V (\varphi))^\perp+ r \hat t.
\end{equation}
Here $G^\perp$ is a matrix determined on the fly to approximate the
inverse of the Hessian in the perpendicular hyperplane; the
approximation is based on the past history of $\varphi$ and does not
require to actually compute the Hessian (for details, see
\cite{ereva01}). In figure~\ref{fig:sevenatomconv}, we compare the
convergence history of the steepest descent method and the
Broyden-accelerated method applied to the seven-atom cluster problem.
The Broyden-accelerated method converges much faster.

\begin{figure*}
  \center \includegraphics{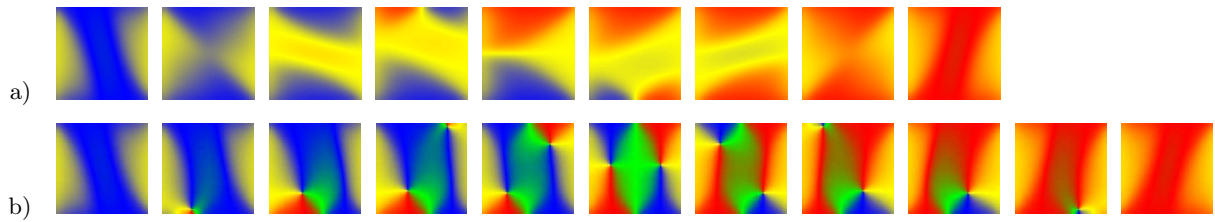}
  \caption{
    Two of the paths [(a) and (b)] followed by the magnetization
    vector $m$ during a switching.  The pictures show the succession
    minimum -- saddle -- ... -- saddle -- minimum. The out-of-plane
    component of $m$ is very small (less than $10^{-2}$) during the
    switching and we only plot its in-plane component with color
    coding: blue = right, red = left, yellow = up, green = down. For
    both paths, we used $N=200$ discretization points along the
    string.} 
  \label{fig:LL1}
\end{figure*}

Once the MEP $\varphi^\star(\alpha)$ has been determined using
Broyden-accelerated string method, free energy barriers and transition
rates can be computed by standard umbrella sampling of the equilibrium
distribution of the system in $S^\star(\alpha)$, the hyperplane normal
to $\varphi^\star(\alpha)$. Consider first the free energy difference
along the string defined as $F(\alpha)-F(0) = -k_B T \ln
(Z(\alpha)/Z(0)$) where
\begin{equation}
  \label{eq:Zdef}
  Z(\alpha) = \int_{S^\star(\alpha)} e^{-\beta V(q)}d^dq
\end{equation}
is the partition function and $\beta = 1/k_B T$.  Using the identity
$\int \partial \ln Z /\partial \alpha d \alpha = \ln Z(\alpha)$, we
obtain from (\ref{eq:Zdef})
\begin{equation}
  \label{eq:freeenergy}
  F(\alpha)-F(0) = \int_0^\alpha \bigl\< (\hat t^\star \cdot\nabla V) 
    \left((\hat t^\star \cdot \varphi^\star)_{\alpha'}
      -\hat t^\star _\alpha \cdot q\right)\bigr\>
  d\alpha'.
\end{equation}
Here $\<\cdot\>$ is the ensemble average with respect to equilibrium
distribution restricted in the hyperplane $S(\alpha)$ and $t^\star$ is
the unit tangent vector along $\varphi^\star$. (\ref{eq:freeenergy})
is similar to the standard thermodynamic integration \cite{frsm96} but
is better suited for numerical purposes. In practice, we use
ergodicity and replace the ensemble average in (\ref{eq:freeenergy})
by a time-average over the solution of an equation similar to
(\ref{eq:SDEV}) but restricted in the hyperplane $S(\alpha)$.
Following Kramers original argument (see e.g.  chap.~9 in
\cite{gar85}), the transition rate can be expressed in terms of the
free energy as
\begin{equation}
  \label{eq:rate}
  k_{A\to B} = \frac{2\sqrt{\lambda_m |\lambda_s|} }
  {\pi \bigl(\gamma +\sqrt{\gamma^2+4|\lambda_s|}\bigr)} 
  e^{-\beta\mathit{\Delta} F},
\end{equation}
where $\mathit{\Delta} F$ is free energy barrier along $\varphi^\star$
\begin{equation}
  \label{eq:fenergyb}
  \mathit{\Delta} F = \max_{0\le\alpha\le 1} \left( F(\alpha) -F(0)\right), 
\end{equation}
and $\lambda_m$ and $\lambda_s$ are (inverse square of) characteristic
time scales at the minimum and maximum of the free energy along the
transition path; $\lambda_m$ and $\lambda_s$ are given by
$|\varphi_\alpha|^{-2} F_{\alpha\alpha}$ evaluated at $\alpha=0$ and
$\alpha=\alpha_s$, respectively, where $\alpha_s$ is the value at
which the maximum in (\ref{eq:fenergyb}) is attained
\cite{remfreeenergy}.  

The transition rates along the MEP obtained earlier for the seven-atom
cluster were evaluated by the string method, and are summarized in
table~\ref{tab:trsevenatom}.

\begin{table}[b]
  \center
  \begin{tabular}{|c||c|c|c|} 
    \hline 
    \hline
    &$k_{A\to B}=k_{D\to C}$ & $k_{B\to A}=k_{C\to D}$ & 
    $k_{B\to C}=k_{C\to B}$ \\ 
    \hline 
    string & $5.023\times10^{-13}$ & $1.425\times10^{-4}$ & 
    $1.211\times10^{-6}$ \\
    \hline 
    exact &  $4.969\times10^{-13}$ & $1.423\times10^{-4}$ &
    $1.206\times10^{-6}$ \\
    \hline
    \hline
  \end{tabular}
  \caption{\label{tab:trsevenatom} 
    The  rates for the various subprocesses in the transition shown in 
    figure~\ref{fig:sevenatom} in the seven-atom cluster problem. 
    We use natural units  and the same parameters as in 
    \cite{deboch98} (for which, e.g.,
    $k_B T/{\mathit \Delta} E_{A\to B}= 0.033$ and 
    $\gamma/2\sqrt{|\lambda_s|}= 0.012$ --low friction limit). 
    The rates  $k_{A\to B}$, $k_{B\to A}$, and $k_{B\to C}$ correspond 
    respectively to the rates for the subprocesses
    $C_0^0 \to C_1^4$, $C_1^4\to C_0^0$, and 
    $C_1^4 \to C_1^3$ identified in \cite{deboch98}. The  values labeled 
    ``string'' were obtained by the noisy string method 
    using (\ref{eq:freeenergy}), (\ref{eq:fenergyb}), and (\ref{eq:rate}). 
    The values labeled ``exact'' were obtained using  
    (\ref{eq:rate}) and (\ref{eq:identity}), 
    by identifying minima and saddle points along the transition path, 
    computing the corresponding 
    energy barrier $\mathit{\Delta} E$,  and evaluating all the eigenvalues 
    of the Hessian at the minima and the saddle points from the Hessian 
    itself.}
\end{table}

The string method can easily be generalized to infinite dimensional
dynamical systems by introducing an appropriate norm in phase-space.
As an example, we consider the problem of thermally induced switching
of a magnetic film. This problem is of great current interest in the
magnetic recording industry \cite{cow00}. (For an introduction to
micromagnetism, see e.g.  \cite{aha96,cow00}; thermally induced
switching is studied in \cite{MAGN}).  Landau-Lifshitz theory of
micromagnetism provides an energy for a ferromagnetic sample $\Omega$
which after suitable nondimensionalization, reads
\begin{equation}
  \label{eq:LLenergy}
  E[m] = A \int_\Omega |\nabla m|^2 d^3x+  \int _\Omega \phi(m) d^3x+
  \int_{\R^3} |\nabla u|^2 d^3x,
\end{equation}
where $m$ is the magnetization distribution normalized so that
$|m|=1$.  The three terms represent respectively energies due to
exchange, anisotropy, and stray field.  The potential $u$, defined
everywhere in space, solves $\hbox{div} \left(-\nabla u + m\right)
=0$, where $m$ is extended as $0$ outside $\Omega$.

\begin{figure}
  \center \includegraphics[width=7cm]{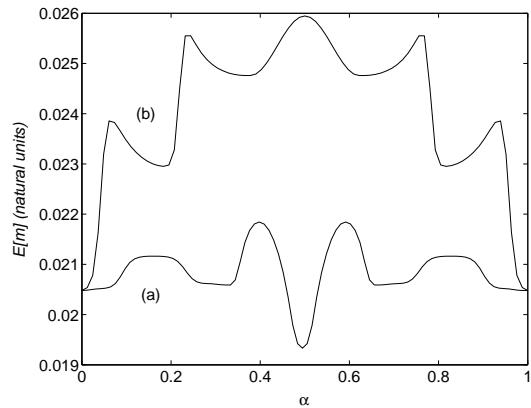}
  \caption{
    The magnetic energy along the two paths [(a) and (b)] shown in
    figure~\ref{fig:LL1} }
  \label{fig:LL2}
\end{figure}

Various switching pathways for (\ref{eq:LLenergy}) were obtained using
the string method: two examples are shown in figure~\ref{fig:LL1} and
the energy along these paths is shown in figure~\ref{fig:LL2}. These
paths illustrate two generic mechanisms for switching in magnetic
films. Path (a), which is more favorable in thin samples, proceeds by
domain wall motion, interior rotation, and switching of the edge
domains.  Path (b), which is more favorable for thicker films,
proceeds by vortex nucleation, invasion of the sample and vortex
expulsion.

In conclusion, transition pathways and transition rates for complex
systems with a relatively smooth energy landscape can be determined
efficiently by evolving strings instead of points in configuration
space. The intrinsic parametrization of the string leads to a simple
and efficient algorithm for the numerical solution of its evolution
equation, and permits to sample the configuration space in regions
that otherwise would be practically inaccessible by standard
Monte-Carlo methods.

We thank Roberto Car and Bob Kohn for helpful discussions.
The work of E is supported in part by NSF grant DMS01-30107.


\begin{thebibliography}{99}
\bibitem{rem_r} In this case $r$ is given explicitly by
%
\begin{displaymath}
  \textstyle
  r = \alpha \int_0^1 \nabla V \cdot \hat t_{\alpha'} d\alpha'
  - \int_0^\alpha \nabla V \cdot \hat t_{\alpha'} d\alpha'.
\end{displaymath}
\bibitem{remfreeenergy} In the small temperature limit, the free
  energy difference can also be expressed as
%
\begin{equation}
  \label{eq:identity}
  \mathit{\Delta} F = \mathit{\Delta} E-k_BT \ln 
  \sqrt{\frac{\hbox{det }H^\perp_m}{\hbox{det }H^\perp_s\ }},
\end{equation}
%
plus higher order terms in $k_BT$, where $\mathit{\Delta} E$ is the
energy barrier along the MEP, and $H^\perp_m$ and $H^\perp_s$ are the
Hessian matrix in the hyperplane perpendicular to the MEP evaluated at
the minimum and the transition state (saddle) respectively. In the
same limit $\lambda_m$ and $\lambda_s$ are the eigenvalues of the
Hessian corresponding to the eigenvectors parallel to the MEP
evaluated at the minimum and the transition state respectively.
\end{thebibliography}
\end{document}